\documentclass[twocolumn]{aastex6}

\usepackage{lineno}

\usepackage{verbatim}
\usepackage{aas_macros}
\usepackage{hyperref}
\usepackage{natbib}
\usepackage{graphicx}
\usepackage{epsfig}
\usepackage{amsmath}
\usepackage[english]{babel}

\shorttitle{A detailed MOND Bullet Cluster}
\shortauthors{X. Hernandez}

\begin{document}


\title{A detailed MOND modelling of the Bullet Cluster}


\author{X. Hernandez}
\affil{Universidad Nacional Aut\'onoma de M\'exico, Instituto de Astronom\'ia, A. P. 70-264, 04510, CDMX, M\'exico.}



\begin{abstract}
 {
The colliding galaxy clusters system 1E 0657-56, the "Bullet Cluster", is often presented as a serious objection to modified gravity theories
dispensing with dark matter in general, and in particular, to the MOND formalism. The argument centres on the fact that the baryonic
matter distribution of this system is dominated by the X-ray emitting gas, while the projected surface density required under General Relativity
to explain the observed lensing signal, centres on the observed galaxies. This spatial offset is interpreted as being in  conflict with MOND, naively
assuming that dark matter being absent, the gravitational potential should centre upon the largest mass distribution, the X-ray gas. However, as under
General Relativity, under MOND, the gravitational potential of a system depends upon the volume density and not just the total mass. I here show 
that the surface density which QUMOND predicts will be inferred under General Relativity from the gravitational potential of the "Bullet Cluster",
qualitatively matches the spatial distribution of what General Relativity inferences of lensing observations return. The close-to-point-like galaxies
imply under QUMOND a relatively larger surface density signal than what is expected from the diffuse gas. As is common under MOND modeling of galaxy
clusters, a mass deficit on the baryon content is obtained, in this case by a factor of 1.6. Beyond this, only small offsets on the details with respect to GR
interpretations of lensing observations remain; MOND is hence shown to be no more discrepant with the "Bullet Cluster" than with any other galaxy cluster.
}  
\end{abstract}



\section{Introduction}

In the debate regarding the interpretation of gravitational anomalies at galactic scales and beyond as either evidence for a change in regime
for gravity at low accelerations, or alternatively as indications for the presence of a dominant hypothetical dark matter component, 1E-0657056, the
``Bullet Cluster'', Tucker et al. (1998), has played a prominent role, e.g. Clowe et al. (2004), Angus et al. (2006). This system shows the recent
collision between a large and a small galaxy clusters; the galaxies belonging to each form two relatively unperturbed distributions some 800 kpc apart,
the gas on the other hand, displays the clear effects of a very strong and disruptive hydrodynamical interaction. The gas of the small cluster now lies
completely stripped about 200 kpc away from the centre of the small galaxy distribution in the direction of the larger cluster, and shows the intense glow
of an unequivocal supersonic shock-wave as a distinct wake signature. The gas of the large cluster, while less perturbed, is also off-centred with respect
to the large galaxy distribution, tending towards the position of the small cluster, also glowing in X-rays and puffed-up with respect to what a quiescent
cluster would show.

A crucial third signal is also present, that of gravitationally lensed background galaxies. This lensing signal centres on the two galaxy distributions,
and when modelled under General Relativity (GR), requires the presence of substantially more matter than present in the stellar populations of the
galaxies themselves. The standard interpretation of this last observation, has been to view it as indicative of the presence of a dominant component
of collisionless hypothetical dark matter. Indeed, the offset between the dominant gas distribution and the gravitational lensing signal
has been presented as evidence against modified gravity theories where no dark matter is hypothesised to exist, e.g. Clowe et al. (2006),
Markevitch et al. (2004). This reasoning has been presented in particular, as an objection against the most well-studied such modified gravity
theory in the Newtonian regime, MOND (Milgrom 1983).

The reasoning for such assertions begins by assuming that under any modified gravity scheme where the gravitational potential of baryonic matter is
boosted so as to account for observed galactic rotation curves in the absence of dark matter halos, necessarily, any gravitational anomaly must
be maximal at the location of the dominant baryonic component. The above conclusion is clearly inconsistent with the empirical configuration of
the bullet cluster. However, the initial assumption in the reasoning is false. Just as under GR, gravitational lensing will depend
not only on the amount of matter present, but also, and crucially, upon the density of the mass configuration. Under modified GR variants having low
velocity limits reproducing MOND, the amplitude of a lensing signal depends on details of the baryonic distribution beyond just the total mass present,
e.g. Campigotto et al. (2017), Verwayen et al. (2024).

Since MOND is not a covariant theory, it is not possible to derive any predictions for what lensing signal one should expect from a given
baryonic matter distribution using only MOND. However, MOND does provide precise predictions for the matter distribution which will be inferred
under a GR framework from any inferences derived from the gravitational potential produced by any given baryonic matter distribution. Under the
QUMOND implementation of MOND, Milgrom (2010), for a given baryonic matter distribution, one can calculate the total gravitational potential as the sum of
the standard Newtonian potential, plus the Newtonian potential associated to an auxiliary mathematical field, the details of which, are fully specified
by the baryonic matter distribution itself. This auxiliary field is termed 'phantom' density, and represents the matter anomaly which under a GR
interpretation of a MOND reality, has to be added to the baryons to understand the gravitational potential present. Hence, in the case of spiral galaxies,
this phantom density will simply be the dark matter halo; an auxiliary construct for facilitating mathematical calculation under MOND, a hypothetical
matter distribution under GR.

Therefore, this allows a comparison of the GR lensing inferences to QUMOND predictions for the Bullet Cluster. Given the baryonic matter distribution
present, one can calculate the phantom density predicted by QUMOND, which after adding to the baryonic matter actually present, and a projection onto
the plane of the sky, can be compared vis-\'a-vis the total surface density maps derived from GR lensing analysis. Such a comparison provides a
quantitative {  spatially resolved and detailed} assessment of the consistency (or lack thereof) of MOND and the present configuration of the Bullet Cluster.

In this {\it letter} I perform such a study and present a first detailed QUMOND analysis of the Bullet Cluster. Making use of the latest Bullet Cluster
galaxy member catalogue and adding the X-ray gas component, it is shown that the total mass surface density distribution which MOND predicts will be
inferred within a GR framework for the Bullet Cluster, spatially closely matches the $\kappa$ maps reported by recent lensing analysis from
Rihtar\v{s}i\v{c} et al. (2026), henceforth R26. Even though the X-ray gas dominates over the stellar galaxies by more than an order of magnitude
in mass, the relatively point-like nature of the latter (considered as purely stellar structures) results in highly concentrated phantom density
distributions which locally significantly dominate over the total surface density (phantom plus baryonic) associated to the very broad diffuse X-ray
gas distribution having a length scale of the order of a Mpc. {  It is general} to MOND studies of relaxed clusters, e.g. Akino et al.
(2022), Kelleher \& Lelli (2024), that the baryon content has to be multiplied by a residual missing mass factor of slightly less than 2 to yield observed
kinematics. In {  consistency with these results, in this case an overall baryonic boost factor of 1.6 is required to match GR lensing
inferences. While an overall normalisation offset appears, no significant displacement or configuration inconsistency of the signal is present.}
The possible causes and implications of these results are discussed in Section 5.

Section 2 gives a brief summary of the salient QUMOND results relevant to this study, and gives analytic examples illustrating the density
dependence of the gravitational anomaly predicted by MOND when viewed in terms of a phantom density distribution. Section 3 presents the
set-up of the baryonic QUMOND model of the Bullet Cluster, and Section 4 gives the resulting phantom plus baryonic matter surface density
distribution and compares it to recent inferences of $\kappa$ maps using GR.

\section{QUMOND analytical expectations}

QUMOND, proposed in Milgrom (2010), is an equivalent formulation of MOND which has the advantage of circumventing
the non-linearities of other MOND proposals, through introducing an auxiliary potential which when added to the
standard Newtonian one, yields the full gravitational potential for a given matter distribution. Under QUMOND, for a given
baryonic matter distribution, $\rho_{\mathrm{b}}(\vec{r})$, to which there corresponds the usual Newtonian potential, $\Phi_{N}$,
through $\nabla^{2}\Phi_{N} =4 \pi G \rho_{\mathrm{b}}$, one must calculate an auxiliary potential $\Phi_{\mathrm{p}}$ to obtain the
total potential as $\Phi=\Phi_{N}+\Phi_{\mathrm{p}}$. This extra potential, termed the "phantom" potential, is sourced by a "phantom"
density $\rho_{p}(\vec{r})$ through $\nabla^{2}\Phi_{\mathrm{p}} =4 \pi G \rho_{\mathrm{p}}$, which in turn is given by:

\begin{equation}
  \rho_{\mathrm{p}}=\frac{1}{4 \pi G} \nabla \cdot \left[ \left( \nu \left( \frac{\lvert \nabla \Phi_{N}  \rvert}{a_{0}} \right) -1  \right)
    \nabla \Phi_{N} \right],
\end{equation}

\noindent where $G$ and $a_{0}$ are Newton's constant and the acceleration scale of MOND, $a_{0}=1.2 \times 10^{-10}$m s$^{-2}$, respectively,
and $\nu(x)$ is a transition function. This 'phantom' density distribution is of course not assumed to be real in any sense, and is nothing more than
a convenient computational auxiliary function. In the high acceleration regime, $a>>a_{0}$, $x>>1$, $\nu(x) \to 1$, the phantom density goes to zero,
and we recover standard Newtonian dynamics. In the low acceleration $a<<a_{0}$ regime, $x<<1$ and $\nu(x) \to x^{-1/2}$ leads to the deep MOND
regime. The details of the $\nu(x)$ transition function are not given by the theory, which only fixes the asymptotic values given above. Many
such functions have been proposed, here we will keep a standard option consistent with the radial acceleration relation across a wide range
of transition values empirically constrained by spiral galactic rotation curve analysis, e.g. Desmond et al. (2024):

\begin{equation}
\nu(x)=\frac{1}{2} \left[1+\left(1+4/x \right)^{1/2} \right].  
\end{equation}

Thus, a given baryonic density $\rho_{\mathrm{b}}$, implies that if a MOND reality is interpreted under a Newtonian framework, the addition
of an extra fictitious matter distribution $\rho_{\mathrm{p}}$ will be needed to reconcile observational inferences of the total gravitational
potential present with observations. From the point of view of MOND, it is this phantom matter density what under standard gravity
is identified as dark matter.

Although the details of the $\nu(x)$ transition function make the direct evaluation of $\rho_{\mathrm{b}}$ cumbersome in general cases,
taking the low acceleration limit allows to calculate this phantom matter density for a number of illustrative cases. If we take a
point mass $M_{\mathrm{b}}$ in the $a<<a_{0}$ limit under the assumption of spherical symmetry, one readily obtains:

\begin{equation}
\rho_{\mathrm{p}}(r)=\left( \frac{M_{\mathrm{b}} a_{0}}{G} \right)^{1/2} \frac{1}{4 \pi r^{2}}.
\end{equation}

This is of course the singular isothermal dark matter halo which must be assumed in addition to the observed baryonic mass of galaxies to understand
the flat rotation curves of spiral galaxies and the constant deviation angle lensing observations for galaxies in general. The details of non-point-like
galactic baryonic matter distributions, together with those of the $\nu(x)$ transition function ensure the excellent fits consistently obtained for MOND
when considering real galaxies, e.g. the radial acceleration relation studies of McGaugh et al. (2016). Notice how this 'extra' component quickly becomes
dominant over the baryonic mass $M_{\mathrm{b}}$ which sources it, as the total phantom mass grows linearly with the radial coordinate. Indeed, the ratio of
phantom mass required under a standard gravity interpretation to the actual baryonic mass present grows as:

\begin{equation}
 \frac{M_{\mathrm{p}}(r)}{ M_{\mathrm{b}}}  = \left( \frac{r}{r_{M}} \right),
\end{equation}

\noindent where $r_{M}=(G M_{\mathrm{b}}/a_{0})^{1/2}$ is the MOND radius, the distance from a point mass beyond which the acceleration falls below $a_{0}$.
For a mass of $10^{11} M_{\odot}$, of the order of the stellar masses of the large elliptical galaxies found in the Bullet Cluster, $r_{M}=10.6$ kpc. Thus,
it is certainly true that MOND expects a significant gravitational anomaly centred upon the elliptical galaxies of the Bullet Cluster, this anomaly
is expected to correspond to the standard 'dark halo' structures needed to be assumed about such galaxies under standard gravity in general.
At a distance of 530 kpc, the phantom matter which under a standard gravity interpretation must be added to the galaxy in question, is already 
50 times larger than the baryonic mass actually present. Also, when viewed as a projected mass anomaly, this phantom matter density is expected by MOND to
have sharply peaked surface density profiles falling as:

\begin{equation}
\Sigma_{\mathrm{p}}(R)=\frac{M_{\mathrm{b}}}{\pi r_{M} R}.  
\end{equation}

Thus, MOND expects a substantial mass anomaly sharply peaked about the elliptical galaxies of the Bullet Cluster. It is clear, that given the dominant
distribution of X-ray gas, MOND also expects the appearance of a mass anomaly centred upon this dominant baryonic component, the relative importance
of which however, will be quite different. The above calculation can be repeated for an extended density profile given by:

\begin{equation}
\rho_{\mathrm{b}}(r)=\rho_{0}\left( \frac{r}{r_{s}} \right)^{-n},  
\end{equation}

\noindent a power law density distribution defined by a scale density and a scale radius, $\rho_{0}$ and $r_{s}$, respectively. The corresponding
calculation now yields the phantom volume {  density as:

\begin{equation}
\rho_{\mathrm{p}}=\frac{5-n}{(3-n)^{1/2}} \left( \frac{a_{0} \rho_{0} r_{s}^{n}}{8 \pi G} \right)^{1/2} r^{-(n+1)/2}.  
\end{equation}

We see that this} phantom density pofile now grows slower than in the case of the point mass of equation(3) for all baryonic density profiles shallower than
$\rho_{\mathrm{b}} \propto r^{-3}$. Indeed, for the constant density $n=0$ profile within the core region of say, a $\beta$ X-ray gas profile, {  the
  relation corresponding to equation(4) of the point mass case becomes:

\begin{equation}
 \frac{M_{\mathrm{p}}(r)}{ M_{\mathrm{b}}}  = \left( \frac{3 a_{0}}{4 \pi G \rho_{0}} \right)^{1/2} r^{-1/2}.
\end{equation}

\noindent while in the point mass case this ratio grows linearly with radius, in the constant density case, this ratio actually falls with radius,
showing that the mass anomaly expected by QUMOND about the highly concentrated galaxies of the Bullet Cluster, will be relatively much stronger than
the signal produced by the diffuse X-ray distribution. Indeed, For numbers typical of the X-ray mass distribution in the Bullet Cluster,
$\rho_{0}=2\times10^{5} M_{\odot}$kpc$^{-3}$ and $r_{s}=630$kpc, e.g., the SZ study of the Bullet Cluster by Halverson et al. (2009), the ratio of equation
(8) evaluated at the same 530 kpc of the previous example, comes to only 1.4. 

This is not to say that the total amounts of phantom matter predicted will depend on the degree of graininess of the sourcing baryonic matter,
Ostrogradsky's theorem as a generalisation of Gauss's theorem, implies that for a given total amount of baryonic matter, the same total amount
of phantom matter will result within a given fixed closed surface. However, the signal due to the X-ray gas over the scale of the cluster, will
be much more diffuse than the singular isothermal phantom matter halos associated with the essentially point-like galaxies, and will have projected
scale lengths which will be larger than the 600 kpc or so scale length of the actual X-ray gas distribution itself. When considered in projection,
the most prominent peaks of phantom density - under GR resulting from the interpretation of the lensing signal - are expected to centre on the galaxies.
}

This is highly analogous to what results in GR, where if we have one neutron star close to a diffuse molecular cloud containing 10,000 times more
mass, the lensing signal will not appear associated to the latter but to the former, regardless of the enormous mass difference in the opposite sense.

While the preceding analytical analysis offers only a rough zero-order estimate of the MOND expectations for the Bullet Cluster, as the details of
the complex mass distribution and $\nu(x)$ function will significantly complicate the full analysis, it does indicate that statements assuming that the
gravitational anomaly which MOND expects for this system must necessarily centre on, and be driven by, the dominant X-ray gas, are naive.  In the
following section I present a detailed QUMOND numerical analysis of the Bullet Cluster using the latest inferences for all the baryonic components
{  and fully detailed QUMOND calculations including the transition function}, to compare against the most recent total mass surface density
inferences under GR through lensing.


\section{Detailed Modelling}

In order to obtain an accurate quantitative evaluation of the gravitational anomaly which MOND predicts given the baryonic matter distribution
of the Bullet Cluster, this section presents a detailed calculation of the phantom matter density which QUMOND predicts directly through equation (1).
This phantom density can then be added to the baryonic matter distribution, and the sum projected onto the plane of the sky to yield the total matter
density which QUMOND predicts will be inferred for this system through a standard gravity interpretation of the potential of the system, i.e., the
total $\kappa$ maps which result from lensing analysis. {  Obtaining an accurate estimate requires a full detailed description of the baryonic
distribution, and an exact theoretical estimate including all the complexities of MOND, beyond the deep MOND limit of the analytical examples described
in the previous section. Hence in this section, the latest observational restrictions will be used for both galaxies and gas, together with a full
QUMOND modelling including a standard transition function an a full 3D modelling of the potential.}

A model cube with the x-axis aligned with the negative R.A. direction with an origin at 104.679167 degrees and the y-axis aligned with the positive
Dec. direction with origin at -55.970278 is set up. The total extent along the x-axis is of 1500 kpc and of 1100 kpc along the y-axis. In order to allow
for an adequate projection of the phantom dark matter along the line of sight, the length of the box covers {  50 Mpc, 25 Mpc in front and 25 Mpc}
behind the plane of the cluster, which places the origin of the z-axis at a distance of 1528.8 Mpc. This prism has a resolution of 750 elements along
the x-axis, 550 along the y-axis and 2900 along the z-axis, for a total of $1.19625 \times 10^{9}$ model elements, giving a resolution of 2 kpc on the
plane of the sky and 12 kpc along the line of sight.

The chosen distance to the origin of the z-axis is what corresponds to the redshift of the Bullet Cluster of $z=$0.296 assuming a flat $\Lambda$CDM
cosmological model with $\Omega_{M}=0.3$, $\Omega_{\Lambda}=0.7$ and $H_{0}=70$ km s$^{-1}$ Mpc$^{-1}$. This is the model used to infer distances
to the galaxies and gas involved in studies using lensing analysis to infer total matter densities, so the same model for inferring distances will be used
here to explore the QUMOND predictions for the total matter density, which under standard gravity should be inferred through observations of the
gravitational potential of the cluster, assuming the same baryonic distribution placed at the same distances.

The first step towards setting up the baryonic mass distribution is the use of the galaxy member catalogue of R26. This catalogue was compiled combining
extant studies of the Bullet Cluster with new spectroscopic observations of several tens of member galaxies, and includes R.A. and Dec. positions,
spectroscopic $z$ and F277W magnitudes for 219 cluster member galaxies, including new JWST observations. In lensing studies of the Bullet
Cluster it is customary to model each galaxy through a PIEMD profile (e.g. Natarajan \& Kneib 1997, Limousin et al. 2005):

\begin{equation}
\rho(r)=\frac{\sigma_{0}^{2}}{2 \pi G}\frac{1}{r^{2}[1+(r/r_{cut})^{2}]},  
\end{equation}

\noindent where $r_{cut}$ is a cutoff radius and $\sigma_{0}$ the central velocity dispersion of the galaxy. The total mass of the above density profile is
$M_{t}=\pi \sigma_{0}^{2}r_{cut}/G$. In the context of lensing modeling, this profile refers to both the stellar component and the dominant assumed dark matter
halo of each galaxy, and is used including a central core radius, $r_{c}$. Here this profile will refer only to the stellar distribution, and hence will be
used in the $r_{c} \to 0$ limit shown above, as the stellar component is closer to a Sersic cusp than to any assumed dark halo core in the centre. Following
R26, this PIEMD profile will be scaled across the galaxies present through the inclusion of standard galaxy cluster scaling laws:

\begin{equation}
  \sigma_{0i}=\sigma_{ref}\left( \frac{L_{i}}{L_{ref}} \right)^{1/4},  r_{cuti}=r_{ref}\left( \frac{L_{i}}{L_{ref}} \right)^{1/2},
\end{equation}

\noindent where $\sigma_{0i}$, $r_{cuti}$ and $L_{i}$ are the density profile parameters and luminosities of each galaxy, and $\sigma_{ref}$, $r_{ref}$ and
$L_{ref}$ are reference values for the velocity dispersion, $r_{cut}$ parameters and the luminosities of the galaxies. Parameters will be chosen such that
$L_{ref}$ corresponds to the F277W magnitude of the brightest cluster galaxy, at $m_{F277W}$=15.77. Since central velocity dispersions are tied to observable
spectral features, $\sigma_{ref}$ will be taken directly from the best fit value found by the lensing analysis of R26 at $\sigma_{ref}$=247 km s$^{-1}$. In
R26 an optimal fit to lensing observations is obtained for $r_{ref}$=101.5 kpc, which puts the mass of a reference galaxy with $m_{F277W}$=15.77 at
4.54$\times 10^{12} M_{\odot}$. This is reasonable if including a dark matter component, as assumed in GR lensing studies, but clearly does not correspond
to the purely stellar components being treated here. {  Thus, I take $r_{ref}$=25 kpc, which brings the assumed total, purely stellar mass of a reference
galaxy with $m_{F277W}$=15.77 at 1.1$\times 10^{12} M_{\odot}$. This mass normalizaion corresponds to taking the mass of the "Bullet" sub-cluster BCG from the
central value used in Clowe et al. (2006) and scaling all galaxy masses through their luminosities as reported in R26.}

The above prescriptions are used to turn
the F277W magnitudes of the 219 galaxies in the Bullet Cluster member catalogue of R26 into individual baryonic density profiles, to be placed at their
corresponding (x, y) coordinates. {  Performing the detailed phantom mass calculation also requires the position of each galaxy along the line of sight.
This last coordinate is taken from a random Gaussian sampling centred at 1528.8 Mpc and having a dispersion of 2Mpc, to account for the range of distances
inferred for the many sub-groups present. Thus, individual distances along the line of sight are not derived from individual galaxy redshifts, as these
mostly reflect internal motions within the cluster.}

The three BCGs are inferred by R26 to be abnormally massive with respect to the overall cluster scaling relations, and will hence be assigned a mass 60\%
larger than what the scaling relations imply, as in that reference. Asides from the individual galaxies, R26 also include four large dark matter halos
with a combined mass of 1.86$\times 10^{15} M_{\odot}$, a component which will be completely excluded from the baryonic QUMOND model.

Regarding galaxies, 7 galaxy groups are also present as described in Benavides et al. (2023) in the vicinity of the Bullet Cluster, groups S1, S2, S3, S5,
S7, S9 and S10 as included also in R26. This list includes the 10 galaxy groups identified in Benavides et al. (2023), and excludes only three which are
already included in the R26 cluster member galaxy catalogue. In this last reference, the velocity dispersion and $r_{cut}$ density profile parameters of
these groups are part of the parameters to be fitted to obtain an optimal reconstruction of the lensing data analysed. Here these groups will be included
as being formed by as many individual galaxies as they each contain (between 6 and 11, see table 2 in Benavides et al. 2023), where each galaxy will be
taken as an average galaxy from the full R26 member catalogue. No extra dark matter halos will be associated to these groups, which will hence be included
with a total mass over an order of magnitude below what results from the lensing fitting of R26. The total resulting stellar mass from all combined galaxies
within the model prism is of 1.696$\times 10^{13} M_{\odot}$.

{  Next}, the X-ray gas will be added through an effective $\beta$ profile distribution as used for the SZ modelling of the Bullet Cluster of 
Halverson et al. (2009), with a centre at RA 06$^h$58$^m$30$^s$.86 and Dec. -55$^{\circ}$ 56' 46.''2, corresponding to  x=449.96, y=383.05 and z=1528800.0
in model coordinates, towards the right edge of the large cluster component. For this distribution a value of $\beta$=1 will be used, consistent with the
value of $\beta=1.15 \pm 0.13$ reported in Halverson et al. (2009) and with the value of $\beta=1.04^{+0.16}_{-0.10}$ inferred by Ota \& Mitsuda (2004) for
this parameter for effective $\beta$ distributions for the X-ray gas of the Bullet Cluster. {  The scale length of this X-ray gas will be of $r_{X}=626.7$
kpc, the central inferred value of this parameter from Halverson et al. (2009) of $r_{X}=626.65 \pm 79.4$}. The central inferred electron density for this
gas will also be taken directly as the central value reported by Halverson et al. (2009), who give $n_{e0}=7.2 \pm 0.3 \times 10^{-3}$cm$^{-3}$. Assuming
$\mu_{e}=1.17$ yields the final X-ray gas profile used as:

\begin{equation}
\rho_{X}(r)=\rho_{X0}\left( 1+(r/r_{X})^{2} \right)^{-1.5},
\end{equation}

\noindent with $\rho_{X0}=2.9\times 10^{5} M_{\odot}$kpc$^{-3}$, resulting in a total X-ray gas mas within the model prism of 25.45$\times10^{13} M_{\odot}$,
15 times larger than the full stellar mass of all the galaxies included. {  Global mass scalings in galaxy clusters yield stellar mass fractions which
decrease towards the largest systems, reaching values compatible with the above quoted ratio at the largest cluster scales, e.g.  Lagan\'a et al. (2013),
Gonzalez et al. (2013). More recently, Akino et al. (2022) have shown that close to a mass scale of $10^{15} M_{\odot}$, of the order of the total
for the Bullet Cluster, X-ray masses comparable to 20 times larger than the stellar masses are found, from a recent sample of 136 X-ray selected galaxy
groups and clusters.}

{ 
  A final minor component to include is the intracluster light, henceforth ICL. This is far less well studied than the other baryonic components and no detailed
  3D modelling of it exists. There is however consensus that the ICL is made up of stars stripped from galaxies, a process which in a highly dynamical
  context such as the collision which gave rise to the Bullet Cluster, is expected to be more relevant than in more common relaxed clusters. The
  total mass of this component in galaxy clusters has been estimated at around 7\% to 23\% of the total stellar galactic mass for merging clusters
  e.g. (Jim\'enez-Teja et al. 2018). Here I will take
  a value for this ratio of 20 \%. The ICL map of the Bullet Cluster from Cha et al. (2025) will be used to model the surface density
  of the gas using two elliptical profiles, one centred on the mid-point of BCG1 and BCG2 with a semi-major axis aligned with the direction defined by these
  two BCGs and measuring 485 kpc and an axis ratio of 2.16. The second ellipse is centred on BCG3 with a semi-major axis of 190 kpc and an axis ratio of
  1.7. The surface density profile of both of these ellipses was taken as:

  \begin{equation}
  \Sigma(R)=\frac{\Sigma_{0}}{R}\left(1-\frac{R}{R_{M}}  \right)^{2},  
  \end{equation}

  Valid out to $R=R_{M}$, where $R$ is a dimensionless distance given by:

  \begin{equation}
    \begin{split}
    R^{2}=\left[\frac{(x-x_{0})\cos(\theta)+(y-y_{0})\sin(\theta)}{a} \right]^{2} \\
    +\left[ \frac{(x_{0}-x)\sin(\theta)+(y-y_{0})\cos(\theta)}{b} \right]^{2}.
    \end{split}
  \end{equation}

  The normalisation constants of both surface densities where chosen so that
  the total mass of the ICL is equal to 20\% of the stellar galactic mas previously calculated, with 25\% of this assigned to the small sub-cluster
  ellipse, and the rest to the main cluster ellipse and $R_{M}$ = 7.45, 2.9 for the main cluster and the sub-cluster, respectively. To avoid the central
  divergence, interior to R=0.5, R was left fixed at 0.5.

  Since there is no 3D modelling for the ICL, this last minority component will not be included in the QUMOND modelling, and the phantom mass associated to it
  will be taken as being the same fraction of its baryonic mass (with the same distribution as the ICL) as results for the other diffuse component,
  that of the gas. Results and conclusions are highly robust to the details of the ICL distribution and it's associated phantom mass, given the very small
  mass fraction involved.
  
}

Once the full baryonic distribution is set up as detailed above, the Newtonian force (as a vector) due to the full baryonic distribution was evaluated at the
centres of each of the $1.19625 \times 10^{9}$ model elements, and used to calculate the $\nu(x)$ QUMOND transition function at each of these points through
equation (2). Then, using a 7 point centred derivative, the QUMOND phantom density was calculated at each of these points. This resulting $\rho_{\mathrm{p}}(r)$
density was then projected along the line of sight.

\begin{figure*}
    
\includegraphics[width=0.525\textwidth, height=210pt]{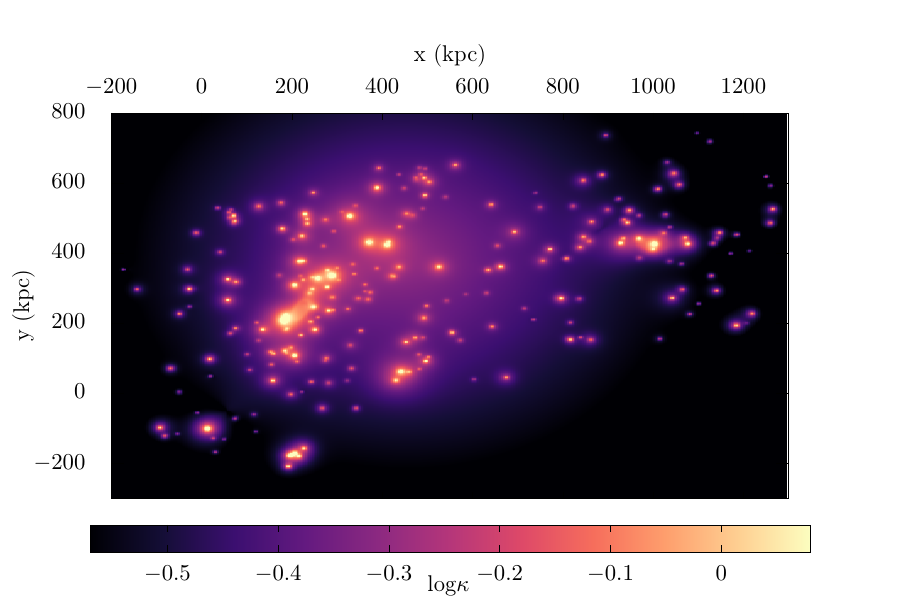}
\hskip -28pt \includegraphics[width=0.525\textwidth, height=210pt]{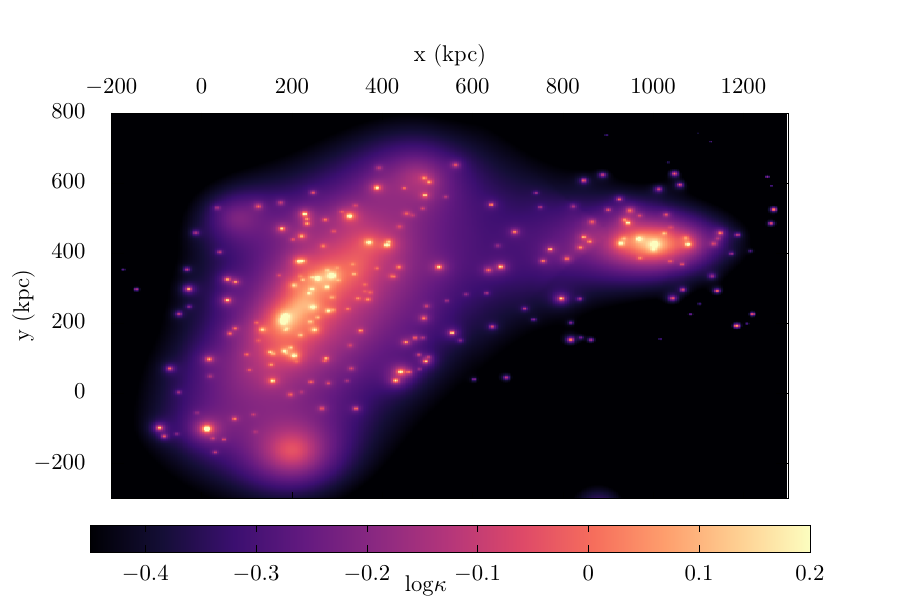}
\caption{Left: This panel shows the total mass surface density projected onto the plane of the sky which QUMOND predicts will be inferred
  when interpreting the baryonic distribution of galaxies and X-ray gas of the Bullet Cluster within the framework of GR. Right:Total mass surface density
  on the plane of the sky as recently inferred by Rihtar\v{s}i\v{c} et al. (2026) for the Bullet Cluster through GR lensing
  modelling. Both figures give mass surface density values in the same units of 1.826876$\times10^{9}$ M$_{\sun}$kpc$^{-2}$.}
\end{figure*}

The full numerical scheme was first tested thoroughly using analytical solutions to the $a<<a_{0}$ limit of
equation (1) for various spherically symmetrical baryonic matter profiles, and checked through both the resulting volume phantom density distributions and
phantom surface density ones. Under MOND modeling of galaxy clusters the ambient external field effect is usually ignored, e.g. Famaey et al. (2025), as the
expected acceleration field external to the cluster is typically of order $a_{ext}/a_{0} \sim 10^{-3}$, e.g. Kelleher \& Lelli (2024), and hence only of
marginal relevance. Therefore, no acceleration field external to the entire system will be included. The external field effect within the Bullet Cluster due
to the complex multi-group configuration will be calculated exactly, through the detailed evaluation of equation (1) including each of the individual galaxies,
the X-ray gas present and all the resulting interactions of the potentials, as already mentioned.

Lastly, to the resulting phantom mass distribution, the full distribution of baryons was then added to obtain a final matter surface density which can
be directly compared to the one inferred under GR lensing studies, as shown in the following section.

\begin{table}
\caption{Total Masses inferred.}
\label{tbl-fig4}
\begin{tabular}{lllllll}
\hline		
                 & $M_{X}$ &  $M_{Gal}$  & $M_{ICL}$ & $M_{\mathrm{p}}$ & $M_{QM}$ & $M_{GR}$              \\		
\hline
Integral         &           &            &          &          &           &                       \\
values           &  2.545    &   0.169    & 0.034    &  7.240   &   9.989   &    12.922             \\ 
obtained         &           &            &          &          &           &                       \\
\hline																	
\end{tabular}

{  The table shows the total baryonic masses in the model volume for gas, galaxies and ICL, first
three entries. The fourth entry gives the total phantom mass which QUMOND predicts. The last two entries show the total mass present in the model volume
as the sum of the baryonic and QUMOND phantom components,} and the total mass in the model region from the R26 GR lensing study.
All masses in units of $M_{\odot} \times 10^{14}$.

\end{table}

\section{Results}

Before discussing the results of the previous section, it must be noted that the baryonic distribution used is not a fit to a set of observational
constraints, but just a first-order estimate of the baryonic mass present {  through central values found in the recent literature.} While lensing
analysis of the Bullet Cluster treat the distribution
of both baryons and of a dominant hypothetical dark matter component through tens of parameters to be fitted in order to achieve an optimal reconstruction
of the lensing signal, the present exercise is simply to use plausible density parameters for all the baryonic components present. Neither the baryonic
distribution parameters nor the QUMOND parameters of the transition function will be adjusted. Thus, the baryonic parameters assumed will simply be used
to assess whether the QUMOND prediction is for a gravitational anomaly dominated by the location of the X-ray gas, as insisted upon when claiming that the
Bullet Cluster somehow falsifies MOND, or if as suggested by the analytical results of Section 2, the QUMOND phantom density is more likely to {  be
defined by the} distribution of galaxies, as happens with the lensing signal. Further, contrary to the case of GR lensing studies, in MOND {  alone} there
is no empirical measure to compare to directly, as the actual configuration of the lensing signal present can not be calculated using a Newtonian theory.

A first description of the results obtained is presented in Table 1, where the total baryonic mass within the model prism is reported for the {  baryonic}
two components included. As already mentioned in the previous section, the mass of the diffuse X-ray gas present is more than an order of magnitude larger
than that of the combined baryonic galaxies. Table 1 also gives the total QUMOND phantom dark matter obtained as a result of the baryons included, { 
of $7.2 \times 10^{14} M_{\odot}$. Hence, the total mass expected by QUMOND to be inferred under GR is of $9.985 \times 10^{14} M_{\odot}$ as given
in Table 1. This value is comparable to recent estimates of the total mass of the Bullet Cluster inferred through GR lensing studies, e.g. Richard et al.(2021),
Cha et al. (2025). More in detail, the kappa map of R26 yields $12.922 \times 10^{14} M_{\odot}$ for the modelled region, leaving a residual missing mass
of  $2.933 \times 10^{14} M_{\odot}$ for the QUMOND model with respect to the recent results of R26.}

The left panel in Figure (1) now shows the full mass surface density distribution over the modelled region as the sum of both the baryonic and QUMOND phantom
mass distributions, in the units usually used to present this quantity in GR lensing studies, of 1.826876$\times10^{9}$ M$_{\sun}$kpc$^{-2}$. This figure is
hence a $\kappa$ map for the Bullet Cluster as predicted by QUMOND for the baryonic distributions present. We see clearly the dominant and 
highly concentrated phantom mass signal predicted by QUMOND about the galaxies, which appear as almost point-like at the scales involved, the final $\kappa$ map
clearly identifies the positions of the galaxies as the locations with the strongest signal. In spite of the diffuse X-ray gas dominating the baryonic
distribution, its relatively much more constant density profile and extent over Mpc scales, results in a close to constant overall $\kappa$ distribution once
integrated over the line of sight. This figure can be compared to the recent GR $\kappa$ map from R26, displayed in the right panel of Figure 1. It is clear
that the overall qualitative morphology of both $\kappa$ maps are highly consistent. Further, the quantitative match is also good, as evidenced by the
small offset of a factor of only {  0.12 in the logarithmic $\kappa$ scales of both figures. This offset corresponds to the 30\% missing mass of the QUMOND
model with respect to the GR results of R26.}

Some differences remain between the GR lensing reconstruction of R26 and the QUMOND $\kappa$ map calculated here, {  beyond the overall 30\% scaling already
mentioned; the right panel is missing a small number of features linked to a few of the galaxies present. This stems from the lack of a uniform coverage of
the lensing features, which after all are only serendipitously available through the fortuitous presence of background lensed galaxies. Further, some of the
galaxy groups modelled as diffuse dark matter halos in R26 appear as collections of galaxies in the QUMOND modelling, and the large central dark matter halos
included in R26 centred on the BCGs are under-represented in the QUMOND maps.}

Crucially, for the Bullet Cluster, MOND does not predict the gravitational anomaly with respect to GR, when vied as an excess mass distribution over the baryons
present, to centre on the gas, but as observed, on the galaxies. {   If the baryonic matter were to be artificially boosted by a constant factor
so as to obtain a total QUMOND mass equal to that reported in R26, this boost factor would be of 1.62, in line with the factor of slightly below 2
generally found for galactic clusters in MOND, e.g. Famaey (2025), Famaey (2026).}

A more detailed quantitative comparison between the $\kappa$ maps of Figure (1) is presented in a $\kappa$ difference map shown in
Figure (2). Some differences remain mostly centred upon the small galaxy groups as mentioned above, e.g. the one close to (200,-200). As the included galaxy
groups were modelled as individual galaxies rather than the assumed dark matter halos of R26, the smooth circular signature of several of these groups are
replaced by the distinct point-like signal of the actual galaxies present in the QUMOND map.

\begin{figure}
\centering
\includegraphics[width=0.525\textwidth, height=210pt]{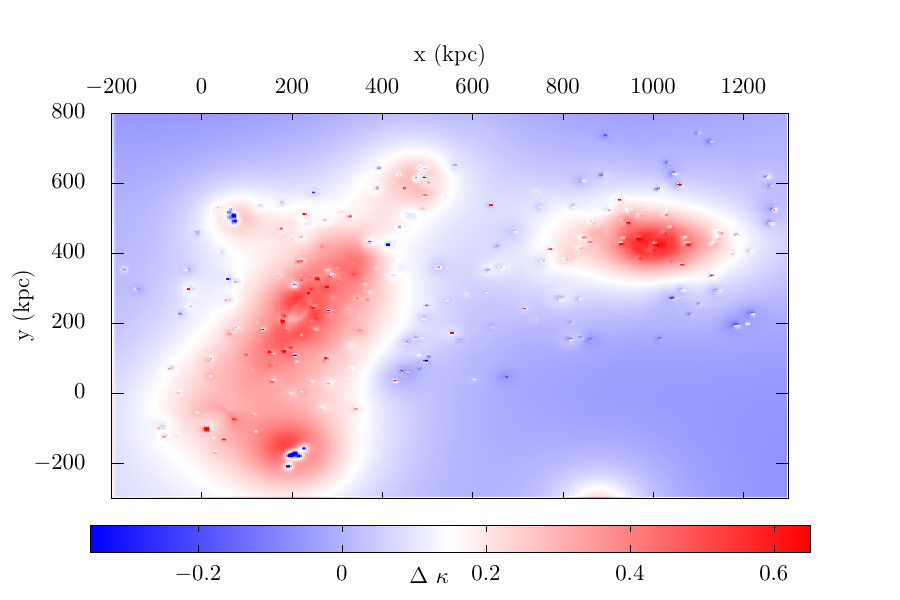}
\caption{Difference map showing the $\kappa$ values reported by Rihtar\v{s}i\v{c} et al. (2026) minus those calculated here. Units of $\kappa$ are of
  1.826876$\times10^{9}$ M$_{\sun}$kpc$^{-2}$.}
\end{figure}

{  Residual differences after the overall 30\% scaling present} mostly fall below 0.15 in the units shown. This level of differences is comparable to
what results from comparing amongst various recent independent studies all using GR lensing analysis, see e.g., the difference maps in figure (7) in R26,
where differences larger than 0.2 are clear. Note also that the differences between independent GR lensing studies have a component associated to the
different assumptions regarding the large-scale behaviour of the hypothetical dark matter halos included, as recently shown by Lin et al. (2023) who
demonstrate that outside of the region of strong lensing signal, the choice of middle and large scale slopes for the dark matter halos used has a determinant
influence on the final inferred matter content.

Through a number of model variations, it was checked that the results presented here are robust, to better than the percentage level, regarding details of the
numerical scheme implemented and small variations of the QUMOND interpolation function within standard ranges.

A final comment regarding the compatibility of the Bullet Cluster with standard $\Lambda$CDM expectations is also relevant. Given the inferred encounter velocity
of the two components as constrained by the supersonic wake evident in the X-ray gas of the small cluster, a number of studies have flagged this system as
inconsistent with standard structure formation scenarios, e.g. Lee \& Komatsu (2010). The details are contentious, with Thompson et al. (2015) giving the number
of expected analogous systems within the mass and relative velocity constraints and the redshift of the Bullet Cluster under $\Lambda$CDM as 1, while after
considering details of the collision and observed configuration, Kraljic \& Sarkar (2015) revise this number to 0.1. Regardless of the details within a
$\Lambda$CDM framework, it is clear that such systems are much more readily created under MOND schemes, e.g. Asencio et al. (2021). This is a relevant point,
as the number of similar objects detected increases e.g., Asencio et al. (2023).

\section{Discussion}\label{ccl}

{ 
A detailed MOND modelling of the Bullet Cluster including individual galaxies as such, had not been performed before. In the absence of any such model,
it was often presented as a fact that any mass anomaly with respect to the baryons would be required by MOND to be defined by the dominant smooth distribution
of X-ray gas. This has now been shown not to be the case, as a QUMOND modelling of the baryonic components yields a resulting $\kappa$ map expected under
GR lensing analysis, which indeed is dominated by the sharply peaked phantom matter density distributions expected about individual galaxies. {  MOND is
known to yield mass deficits when modelling galaxy clusters, as indeed obtained here. However, no dominance in the expected total $\kappa$ map of the
X-ray distribution appears, showing the Bullet Cluster not to be an any more serious objection to MOND than any other galaxy cluster.}

The GR and QUMOND $\kappa$ maps, although similar, are not identical. A residual missing mass {  of $2.937 \times 10^{14} M_{\odot}$ remains} for the QUMOND
model with respect to the R26 results. This number is a lower limit, as it is restricted to the region modelled, including a larger aperture would slightly
increase it. Also, the lensing coverage is not complete, as evidenced by the presence of a few distinct peaks in the QUMOND map which have no counterpart
in the R26 $\kappa$ map, about member galaxies which simply do not happen to have a background lensed object.

Results presented here are comparable to recent MOND modelling of relaxed galaxy clusters, where the galaxies are treated as averaged spherical distributions,
e.g. Famaey et al. (2025) or McGaugh et al. (2026), where comparable missing masses are reported. {  In particular, the first order assessment of the Bullet
Cluster phantom mass distribution of Famaey (2026), which appeared during the revision process of this paper, quotes $3.4 \times 10^{14} M_{\odot}$ as the
missing mass under QUMOND for this system.}

{  Beyond the 30\% offset in total amounts of matter inferred, which would correspond to boosting the Baryonic component by a factor of 1.62,} the QUMOND map
does reveal a deficit about the central cluster-scale dark matter halos assumed under GR. This appear mostly within the regions of strong lensing close to the
BCGs, and could hence reveal a small failure of QUMOND in the details at such high velocity dispersion systems, perhaps indicative of the appearance of the
first relativistic corrections to MOND, or perhaps the presence of some small amounts of an extra dissipationless component as neutrinos, as it has also
sometimes been suggested, Famaey \& McGaugh (2012).

Performing similar detailed and comparative studies of other collision galaxy clusters and relaxed ones, finally treating these systems as collections
of individual galaxies, will help towards a more complete assesment of the behaviour of gravity at the low acceleration scale, and allow to better inform
the development of future theoretical developments.
}

\section*{Acknowledgements}

I acknowledge the constructive input of Benoit Famaey and of another anonymous referee as crucial towards helping to clarify my
original manuscript and clear up an initial mistake.
I also acknowledge the kind help of Gregor Rihtar\v{s}i\v{c} and Randall Scottt for assistance in obtaining the results of the lensing
modelling reported in Rihtar\v{s}i\v{c} et al. (2026), of Sangjun Cha with use and handling of the ICL observations of Cha et al. (2025)
and of H\'{e}ctor Ibarra Medel in handling .fits files. X.H. acknowledges financial assistance from SECIHTI SNII and UNAM DGAPA PAPIIT
grant IN-102624.

\end{document}